\documentclass[twocolumn,aps,prb]{revtex4-1}
\pdfoutput=1 

\usepackage[T1]{fontenc}
\usepackage{lmodern}
\usepackage{txfonts}

\usepackage{fancyhdr}
\fancypagestyle{append}{%
        \fancyhead[LE,RO]{\thepage}
        \fancyhead[LO,Re]{\textit{YBCO thin films},
         ~~~M.W. Denhoff and J.P. McCaffrey}
          \fancyfoot[C]{Appendix, Apr 1991}
           }
\fancypagestyle{plain}{%
        \fancyhead[LE,RO]{}
        \fancyhead[LO,Re]{}
          \fancyfoot[C]{Preprint, Apr 1991}
           }

\usepackage{graphicx}
\usepackage{dcolumn}
\usepackage{bm}
\usepackage{textcomp}

\graphicspath{{pic/}}

\usepackage{listings}
\usepackage{verbatim}

\usepackage[colorlinks=true, allcolors=blue]{hyperref}

\begin{document}



\title{Epitaxial Y$_1$Ba$_2$Cu$_3$O$_7$ thin films on CeO$_2$ buffer layers
on sapphire substrates}

\author{M. W. Denhoff}
\author{J. P. McCaffrey}
\affiliation{National Research Council of Canada, 1200 Montreal Road,
	    M-50, K1A 0R6, Ottawa, Ontario, Canada\vspace{1ex} \\
            This is the ``accepted manuscript'' of the article published as \\
M. W. Denhoff and J. P. McCaffrey, ``Epitaxial Y$_1$Ba$_2$Cu$_3$O$_7$ thin
films on CeO$_2$ buffer layers on sapphire substrates'',  J. Appl. Phys.
\textbf{70} (7), 1 October 1991}
\date{April 1991}
\email{mwdenhoff@gmail.com}



\begin{abstract}
Pulsed laser deposition has been used to deposit Y$_1$Ba$_2$Cu$_3$0$_7$
layer on CeO$_2$ buffer layers on (1\underline{1}02) sapphire.
Both layers are epitaxial 
with the (110) direction of the CeO$_2$ layer aligned with the
$<$\underline{2}021$>$ direction of the sapphire substrate. The c-axis
Y$_1$Ba$_2$Cu$_3$O$_7$ layer has its $<$100$>$ direction
aligned with the $<$110$>$ direction of the CeO$_2$. Cross-sectional
transmission electron microscopy shows the epitaxy to be coherent and the
interfaces to be abrupt at an atomic level. The best films have a critical
current of $9\times 10^6$\,A/cm$^2$ at 4.2\,K and lower microwave surface
resistance than copper at 77 K and at a frequency of 31 GHz.
\end{abstract}

\maketitle
\thispagestyle{plain}


Electronic applications of high-temperature superconductor
(HTS) thin films require films with high critical
currents and uniform properties over the entire surface of
the substrate. Due to the extreme anisotropy and short
coherence lengths of HTS materials, thin films must be
epitaxial in order to have the above properties. Good quality
HTS films have been grown on substrates such as
SrTiO$_3$, MgO, yttria stabilized zirconia, and LaA1O$_3$.
These substrates have various shortcomings for electronic
applications; high microwave losses, high dielectric constants,
low mechanical strength, high cost, and lack of
availability of large good quality single crystals. On the
other hand, Si and sapphire (Al$_2$O$_3$) are excellent electronic
materials and are commercially available in the form
of essentially perfect, single-crystal wafers with epitaxial
quality surfaces. Unfortunately, Si and Al react chemically
with HTS materials at the thin-film growth temperatures.
Attempts to grow HTS films directly on Si or sapphire use
low growth temperatures and high deposition rates as well
as requiring thick films to get away from the reacted
interface.\cite{Berezin, Boyce}  These techniques are in opposition to the
requirements of epitaxial growth. A possible solution to this
problem is to use a chemical diffusion barrier. The requirements
of such a buffer layer are that it should not react
significantly with the substrate or the HTS film and the
layer must grow epitaxially on the substrate as well as
allow the subsequent HTS layer to be
epitaxial.\cite{Wu, Wiener-Avnear, Talvacchio, Kingston, Wu2}

CeO$_2$ has several properties which make it a good candidate
as a buffer layer. It is a refractory material that is
cubic with a lattice constant of 5.411\,Å which is nearly the
same lattice constant as Si. The dielectric constant of
CeO$_2$ is about 26, we do not know what its microwave
frequency losses are. CeO$_2$ has been grown epitaxially on Si
by electron beam evaporation.\cite{Inoue}  Since Si grows epitaxially
on sapphire, it is likely that CeO$_2$ will do so also. It has also
been determined that CeO$_2$ and Y$_1$Ba$_2$Cu$_3$0$_7$ (YBCO) are
mutually insoluble.\cite{Sampathkumaran, Sladeczek}  There is a good
possibility of obtaining
a very clean interface on the scale of atomic ordering.
We have found that CeO$_2$ will grow epitaxially on MgO
and sapphire and that good quality YBCO can be grown
on the CeO$_2$ buffer layer. In this letter, we will report on
the deposition procedures and on the microstructure of
YBCO on CeO$_2$ films on (1\underline{1}02) oriented sapphire.

     CeO$_2$ layers were deposited by pulsed laser deposition
(PLD) using techniques similar to the PLD methods that
have been reported elsewhere.\cite{Boyce, Wu3}  The depositions were
made in 0.20\,Torr of oxygen, using a 248\,nm HeF eximer
laser focused to a spot giving a fluence of about
3\,J\,cm$^{-2}$. The substrate is mounted, using silver paint, on an
Al$_2$O$_3$ ceramic plate which is heated from behind by a
nickel-chromium resistive heating ribbon. Substrate temperatures,
measured using an optical pyrometer, of about
740\,°C are used for the CeO$_2$ deposition.

     The CeO$_2$ layers used as buffer layers for YBCO were
less than 20\,nm thick and deposited by operating the laser
at 10 Hz for a time of 120\,s. The substrates were cooled in
500 Torr of oxygen and then the chamber let up to air to
allow a YBCO target to be exchanged for the CeO$_2$ target.
A YBCO layer was then grown following the same procedure
as was used for the Ce02 layer except for a slightly
lower substrate temperature of 720\,°C. YBCO layers between
300 and 500\,nm thick were grown over deposition
times of 20 to 35 min.

     In this report we will concentrate on structural analysis,
however we will first present a summary of electrical
properties. The resistance ratio {R(295\,K)/R(100\,K)} is
about 2.8 for these films. The better films have a sharp
transition with zero resistance above $(88\pm 0.5)$\,K, whereas
poorer films have a tail in the resistive transition but still go
to zero resistance above $(84\pm 0.5)$\,K. The normal
resistance of 260\,\textmu\textohm\,cm at 295\,K and
950\,$\textmu\textohm$\,cm at 100 K are
about the same as typical values quoted in the literature for
YBCO. Over 50\% of the films we grow are in the better
category. X-ray diffraction and scanning electron microscopy
(SEM) examination (including electron channeling)
do not reveal a difference between the two. The critical
current density of a 100\,\textmu m-wide line
was $1.3\times 1^{-6}$\,A\,cm$^{-2}$ at 84\,K.

     The microwave surface resistance of the films was measured
qualitatively by placing the films on the end of a
copper cavity immersed in liquid nitrogen. The $Q$ was then
measured at the resonant frequency of 31 GHz. Several
films have given higher $Q$’s than a copper end plate. Our
system is not sensitive enough to quantify the surface
resistance of the better films.

     One sample was examined by a Faraday rotation
magneto-optic technique.\cite{Batalla}  The 1-cm-square, 3200-Å-thick
YBCO film was found to shield an applied field of 9\,mT
over its entire surface at 4.2\,K. The observed trapped field
implied a critical current density of $9\times 10^6$\,Acm$^{-2}$ in a
field of 20\,mT.

\begin{figure} [t]
\begin{center}
\includegraphics[width=0.48\textwidth]{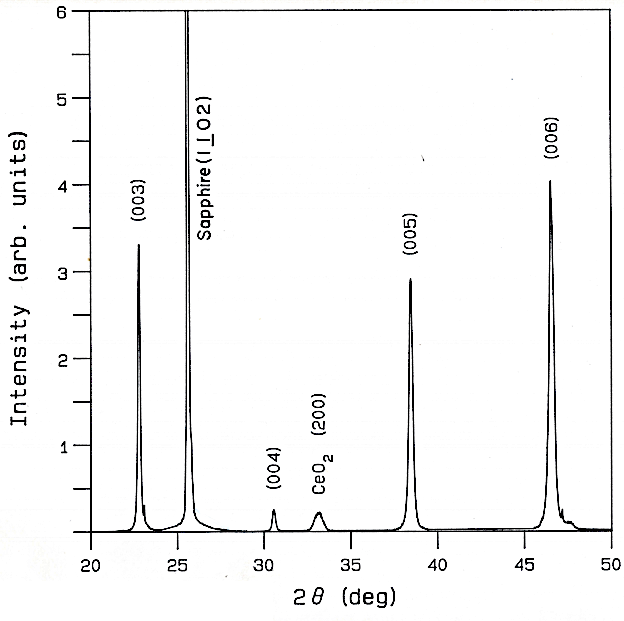}
\end{center}
\caption[example]
{ \label{fig:fig-1}
$\theta$-$2\theta$ x-ray diffraction spectrum using CuK radiation. The YBCO
c-axis lines are indexed.}
\end{figure}

     The YBCO layers are single phase c-axis oriented as
shown by the x-ray diffraction spectrum in Fig.~\ref{fig:fig-1}. The lines
are strong and sharp [the half width of the (005) line is
0.26°] which is consistent with an epitaxial layer. The small
peaks on the high angle side of the (003) and (006) peaks
correspond with the lattice spacing of the (100) and (200)
planes of the tetragonal phase, YBa$_2$Cu$_3$O$_7$. The only
CeO$_2$ line present is the (200) line [we also see the (400)
line at 2$\theta$ = 69.5Å]. This indicates a single orientation of the
Ce0$_2$ layer. The line is broadened having a width
$(\Delta 2\theta$) = 0.56°. If this broadening is due only to the thickness
of the layer, the thickness can be calculated using\cite{James}
$t =\lambda/[(\Delta 2 \theta) \cos(\theta)]$, where $\lambda = 0.15406$\,nm.
This gives t = 16\,nm which is slightly larger than the thickness
measured by TEM.

\begin{figure}[t]
\begin{center}
\includegraphics[width=0.48\textwidth]{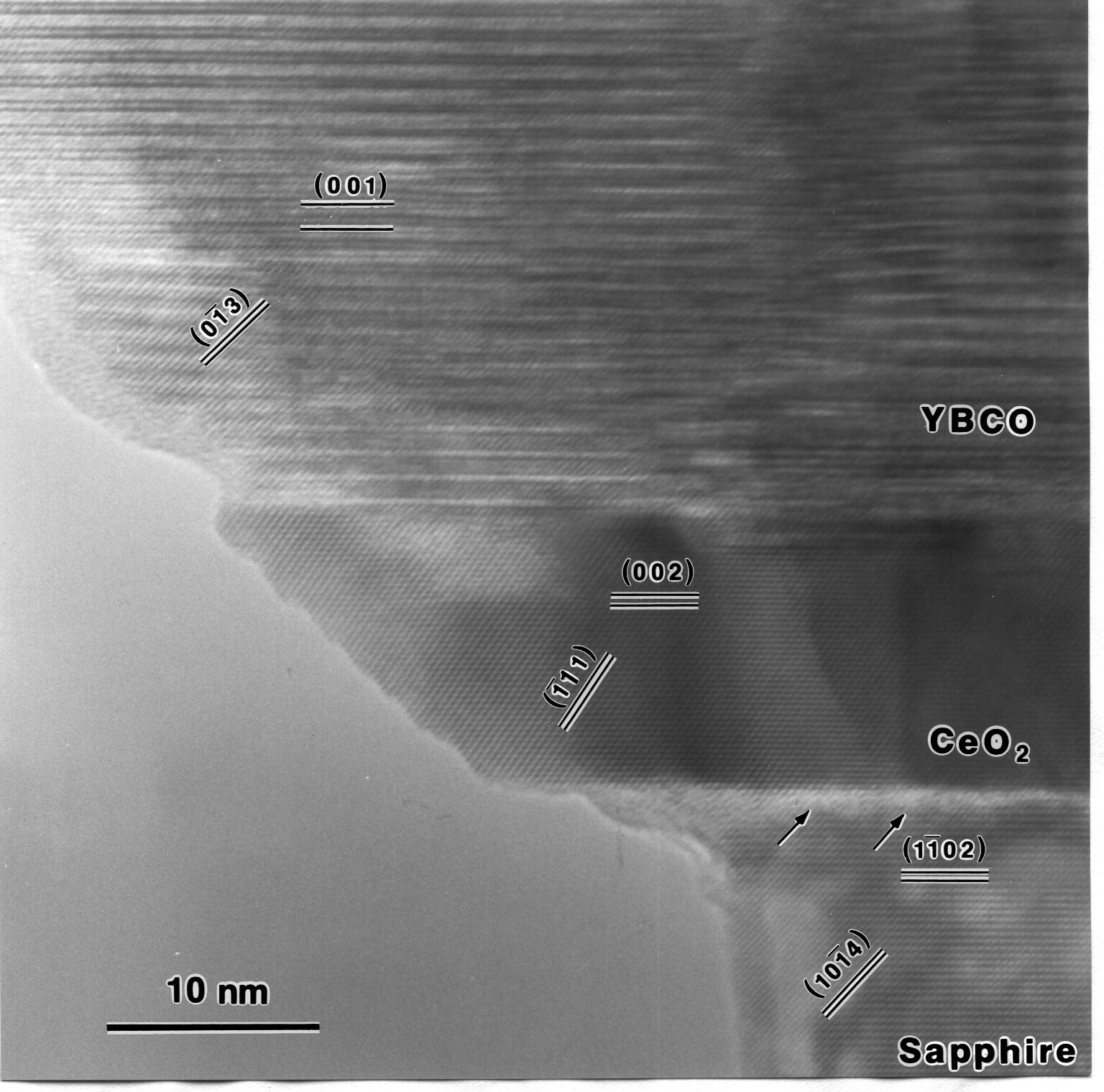}
\end{center}
\caption[example]
{ \label{fig:fig-2}
High-resolution TEM micrograph of YBCO/CeO$_2$/sapphire,
looking down the following zone axes; sapphire $<$\underline{2}021$>$,
CeO$_2$ $<$110$>$, and YBCO $<$100$>$. Two sets of planes are indicated
for each material. The arrows point towards misfit dislocations.}
\end{figure}

     Transmission electron microscopy (TEM) was performed
using a Philips EM430T operating at 250\,kV. Samples
were prepared by a small-angle cleavage technique
described elsewhere.\cite{McCafFrey}  Shown in Fig.~\ref{fig:fig-2}
is a cross section
looking down the sapphire $<$\underline{2}021$>$ zone axis. In agreement
with the XRD results the growth direction of the CeO$_2$ is
$<$001$>$ and the growth direction of the YBCO is $<$001$>$.

     The in-plane alignment is with the (110) zone axis of
the CeO$_2$ parallel with the $<$\underline{2}021$>$ sapphire zone axis. The
same alignment was found by electron channeling measurements
made using a scanning electron microscope.

This is the same epitaxial alignment that is seen for Si on
(1\underline{1}02) sapphire.\cite{Filby} 15 The lattice mismatch is found noting
that the (11\underline{2}0) and the (\underline{1}104) planes in sapphire are
parallel to the (100) and (010) planes of the CeO$_2$. The
spacings of 0.2379 and 0.2552\,nm of the (11\underline{2}0) and the
(\underline{1}104) planes in sapphire have mismatches of 12\% and
5.7\% with the spacings CeO$_2$ (100) planes. Any strain
induced by the epitaxy due to such large lattice mismatches
should be relaxed over the first two or three atomic layers
by the formation of misfit dislocations. This is what we see
in the cross-sectional micrograph. There is a disordered
layer about two atomic rows ($\sim 0.5$\,nm) thick at the
sapphire/Ce0$_2$ interface, which appear lighter in the
micrograph due to strain contrast effects. Although the
interface is disordered a majority of the (10\underline{1}4) sapphire planes
line up with (111) Ce0$_2$ planes with an 8° tilt at the
interface. This shows that the epitaxy is coherent.

     The CeO$_2$ buffer layer in Fig.~\ref{fig:fig-2} is about 13\,nm thick.
The top has atomically flat areas which appear to be separated
by steps that are about 1 to 1.5\,nm high. The epitaxial
in-plane relation is that the (100) direction of the
YBCO is parallel to the $<$110$>$ CeO$_2$ direction. Since the
lattice mismatch for this orientation is less than 1\%, the
strain needed to accommodate the heteroepitaxy is small.
This is consistent with the micrograph as there is no strain
contrast at the YBCO/CeO$_2$ interface. The (111) planes of
the CeO$_2$ are continuous with the (013) planes of the
YBCO with a tilt of about 10° at the interface.

     The epitaxial alignment for these layers, i.e., YBCO
$<$100$>$ parallel to sapphire $<$\underline{2}021$>$, is the same that has
been observed for YBCO grown directly on (1\underline{1}02)
sapphire.\cite{Boyce}  The same alignment has also been reported for
YBCO on an yttria stabilized zirconia buffer layer on
sapphire,\cite{Wu2} in which the buffer layer is oriented in the same
way as the CeO$_2$ reported here. It is interesting that the
YBCO maintains the same alignment with the sapphire
with or without a buffer layer.

     We would like to comment on two further points. The
good lattice match and the insolubility of Ce in YBCO
make CeO$_2$ a very good candidate for the insulating layer
in a tunnel junction. This is due to the short coherence
depth and the sensitivity of the superconducting properties
of YBCO to strain and chemical impurities. Strain damage
or chemical diffusion from the buffer layer traveling only
one or two lattice constants into the YBCO would likely
severely degrade tunneling properties. The results shown in
Fig.~\ref{fig:fig-2} suggest it may be possible to grow a smooth
continuous 2-nm-thick CeO$_2$ insulating layer between two YBCO
layers. The second point is that the lattice fringes of the
CeO$_2$ and sapphire extend right to the edge of the sample
which had been exposed to the atmosphere for several days
before TEM investigation. This is a dramatic demonstration
of how these rugged materials maintain epitaxial quality
surfaces. This is not true for the YBCO layer which is
known from surface analysis methods to quickly develop a
damaged top layer of carbonates and hydroxides.\cite{Budhani} At the
edge of the sample in Fig.~\ref{fig:fig-2}, one can see an amorphous
layer of about 2.5\,nm thickness.

     We have shown that YBCO films with excellent structural
quality can be grown on sapphire with a CeO$_2$ buffer
layer. This is due to the epitaxial CeO$_2$ buffer layer which
is well lattice matched to YBCO and allows the use of a
higher growth temperature than can be used for YBCO
directly grown on sapphire.

     The magneto-optic measurements were done by E.
Batalla, some electrical measurements were done by Suso
Gygax, and the microwave measurements were done by
Dick Clark. We thank Peter Grant for helpful discussions.
Finally, we greatly appreciate the careful work of Hue
Tran who deposited the films.


\end{document}